\documentstyle[12pt]{article}

\begin{document}
\topmargin 0pt
\oddsidemargin 7mm
\headheight 0pt
\topskip 0mm

\addtolength{\baselineskip}{0.40\baselineskip}

\hfill SOGANG-HEP 214/97

\hfill February 20, 1997

\begin{center}

\vspace{36pt}
{\large \bf Superalgebraic Truncations from {\it D}=10, {\it N}=2
Chiral Supergravity}

\end{center}

\vspace{36pt}

\begin{center}

Chang-Ho Kim

{\it Department of Physics, Seonam University, Namwon, Chonbuk 590-170, Korea}

\vspace{12pt}

Young-Jai Park$^*$

{\it Department of Physics and Basic Science Research Institute, \\
Sogang University, C.P.O. Box 1142, Seoul 100-611, Korea}

\end{center}

\vspace{1cm}
\begin{center}
{\bf ABSTRACT}
\end{center}

We study ten-dimensional {\it N}=2 maximal chiral supergravity in the
context of Lie superalgebra SU(8/1).  The possible successive superalgebraic
truncations from ten dimensional {\it N}=2 chiral theory to the 
lower dimensional supergravity theories are
systematically realized as sub-superalgebraic chains of SU(8/1) by using the Kac-Dynkin weight techniques.

\vspace{2cm}


\vspace{12pt}

\noindent

\vspace{12pt}

\vfill
\hrule
\vspace{0.5cm}
\hspace{-0.6cm}$^*$ E-mail address : yjpark@ccs.sogang.ac.kr

\newpage
\begin{center}
{\large \bf I. Introduction}
\end{center}

There have been considerable interests in superalgebras which are relevant to many supersymmetric theories.$^{1,2}$
Recently, M and F theories$^3$ have been  also tackled from the point of view of the general properties of the superalgebra.$^4$
Supersymmetric extensions of Poincar\'{e} algebra in
{\it D}-dimensional space-time were reviewed, and their representations (reps) for the supermultiplets of all known supergravity theories were extensively searched by Strathdee.$^5$ This work has been an extremely useful guideline for studying supersymmet
ric theories.  Cremmer$^6$ developed the complicated method for consistent trunctions by choosing a particular rep of real symplectic metric in order to derive {\it N}=6,4,2 supergravities from {\it N}=8 in five dimensions.

On the other hand, during last ten years, we have shown that superalgebras allow a more systematic analysis for finding the supermultiplets$^{7,8}$ of several supergravity and type-IIB closed superstring theories by using the Kac-Dynkin weight
techniques of SU({\it m}/{\it n}) Lie superalgebra.$^9$
In particular, we have shown that the massless reps of supermultiplets of the {\it D}=10, {\it N}=2 chiral supergravity$^{10}$ and the {\it D}=4, {\it N}=8 supergravity$^{11}$ belong to only one irreduclble representation (irrep) of the SU(8/1) superalgeb
ra using the Kac-Dynkin method.$^{12}$  Recently, we have shown that all possible successive superalgebraic truncations from four-dimensional {\it N}=8 theory to {\it N}=7,6,..,1 supergravity theories are systematically realised as sub-superalgebra chains
 of SU(8/1) superalgebra.$^{13}$

In this letter, we show that the successive superalgebraic trunctions from {\it D}=10, {\it N}=2 chiral supergravity$^9$ to possible lower dimensional nonmaximal theories can be easily realized as sub-superalgebra chains of SU(8/1) Lie superalgebra by usi
ng projection matrices.$^{14}$  In Sec. II, we
briefly recapitulate the mathematical structure of the SU(8/1) superalgebra related to {\it D}=10, {\it N}=2 maximal chiral supergravity.  In Sec. III, we explicitly show that supermultiplets of possible lower dimensional supergravity theories can be syst
ematically obtained from SU(8/1) by successive superalgebraic dimensional reductions and truncations.
The last section contains conclusions.

\vspace{1cm}
\begin{center}
{\large \bf II. Kac-Dynkin Structure of SU(8/1) superalgebra}
\end{center}

In this section, let us briefly recapitulate the Kac-Dynkin Structure of SU(8/1) superalgebra. The Kac-Dynkin diagram of the SU(8/1) Lie superalgebra is

\begin{eqnarray}
w_1~~~~~w_2~~~~ w_3~~~~ w_4~~~~~ w_5~~~~ w_6~~~~ w_7~~~~~ w_8~ \nonumber \\
\bigcirc \!\!-\!\!\!-\!\!\!-\!\!\bigcirc \!\!-\!\!\!-\!\!\!-\!\!
\bigcirc \!\!-\!\!\!-\!\!\!-\!\!\bigcirc \!\!-\!\!\!-\!\!\!-\!\!
\bigcirc \!\!-\!\!\!-\!\!\!-\!\!\bigcirc \!\!-\!\!\!-\!\!\!-\!\!
\bigcirc \!\!-\!\!\!-\!\!\!-\!\! \bigotimes 
\end{eqnarray}

\noindent
where the set $(w_1~w_2~\cdots~w_8)$ characterizes the heighest weight vector of an irrep.$^{1,2}$  The components $w_i~(i \neq 8)$ of this vector should be a nonnative integer, while $w_8$ can be any {\it complex} number. The last node denotes the simple
 odd
root $\beta_8$, while the seven white nodes in the Kac-Dynkin diagram denote the simple even roots
$\alpha_i~(i=1,~2,~\cdots,~7)$, which constitute SU(8) subalgebra.

The corresponding graded Cartan matrix is given by

\vspace{0.5cm}
\begin{equation}
\left [
\begin{array}{rrrrrrrr}
2  & -1 & 0  & 0  & 0  & 0  & 0  & 0  \\
-1 & 2  & -1 & 0  & 0  & 0  & 0  & 0  \\
0  & -1 & 2  & -1 & 0  & 0  & 0  & 0  \\
0  & 0  & -1 & 2  & -1 & 0  & 0  & 0  \\
0  & 0  & 0  & -1 & 2  & -1 & 0  & 0  \\
0  & 0  & 0  & 0  & -1 & 2  & -1 & 0  \\
0  & 0  & 0  & 0  & 0  & -1 & 2  & -1 \\
0  & 0  & 0  & 0  & 0  & 0  & -1 & 0
\end{array}
\right ]
 .
\end{equation}
\vspace{0.5cm}

\noindent
Note that each positive simple even root $\alpha_i^+$ corresponds to the {\it i}-th column of the graded Cartan matrix, while the positive simple odd root $\beta_8^+$ corresponds to the last column of the graded Cartan matrix.  The negative simple roots $
\alpha_i^-$ and $\beta_8^-$ are given by

\begin{equation}
\alpha_i^-  =  - \alpha_i^+,~~    \beta_8^-  =  - \beta_8^+,
\end{equation}

\noindent
and other odd roots are easily obtained by

\begin{equation}
\beta_i^\pm = [\alpha_i\pm ,~ \beta_{i+1}\pm],~~~i=1,~2,~\cdots,~7.
\end{equation}

\noindent
Then, the action by an odd root $\beta_i\pm$ alternates a bosonic (fermionic) floor with a fermionic (bosonic) one.

The fundamental rep of SU(8/1) is (1~0~0~0~0~0~0~0), and it has the substructure 
$[({\bf 8,1})_F ~\oplus~({\bf 1,1})_B]$ in the SU(8)$\otimes$U(1) bosonic subalgebra basis, where the subscripts {\it F} and {\it B} stand for fermionic and bosonic degrees of freedom, respectively, as follows:
\\
\begin{equation}
\begin{array}{lccl}
                 & (1~0~0~0~0~0~0~0) \\
\\
\mid \mbox{ground}> & (1~0~0~0~0~0~0~0) & = & ({\bf 8},~{\bf 1})_F \\
                 & \Downarrow \beta_1^- \\
\mid \mbox{1st}> & (0~0~0~0~0~0~0~1) & = & ({\bf 1},~{\bf 1})_B. \\
\end{array}
\end{equation}
\\
\noindent
On the other hand, the complex conjugate rep of the fundamental rep is given by $(0~0~0~0~0~0~0~-1) =$
$[({\bf 1,1})_B~ \oplus$ $({\bf \overline {8},1})_F]$.
The even and odd roots consist of the adjoint rep $(1~0~0~0~0~0~0~-1)$, which is obtained from the tensor product of the above two reps, as follows
\\
\begin{equation}
\begin{array}{llcl}
                 & (1~0~0~0~0~0~0~-1) & \\
\\
\mid \mbox{gnd}> & (1~0~0~0~0~0~0~-1) & = & \beta_i^+ \\
\\
\mid \mbox{1st}> & (1~0~0~0~0~0~1~-1) & = & \mbox{SU(8)} \\
                 & (0~0~0~0~0~0~0~0) & = & \mbox{U(1)} \\
\\
\mid \mbox{2nd}> & (0~0~0~0~0~0~1~0) & = & \beta_i^-. \\
\end{array}
\end{equation}
In general, there are two types of irreps of SU({\it m}/{\it n}),
which are {\it typical} and {\it atypical}.$^{1,9,15}$  All atypical reps of SU(8/1) are characterized by the last component of the highest weight.  The atypicality condition$^9$ is given by

\begin{equation}
w_8 = - \sum_{j=i}^{7} w_j + i - 8,~~~1\leq i\leq 8.
\label{line1}
\end{equation}

\noindent
Note that since an odd root $\beta_i^-$ string is terminated in the full weight system for the case of the atypical rep such that $w_8$ satisfies Eq.(7) for a specific {\it i}, the atypical reps generally have not equal bosonic and fermionic degrees of fr
eedom.

On the other hand, all the typical reps of SU(8/1) consist of nine floors and have equal bosonic and fermionic degrees of freedom.  The typical, lowest dimensional rep is
$(0~0~0~0~0~0~0~w_8)~=$ $[{\bf 128}_B \oplus {\bf 128}_F]$ $= [{\bf 1} \oplus {\bf \overline 8} \oplus {\bf \overline {28}} \oplus
{\bf \overline {56}} \oplus  {\bf 70} \oplus {\bf 56} \oplus {\bf 28}
\oplus {\bf 8} \oplus  {\bf 1}]$ for
$w_8 \neq$ $0,~-1,~\cdots,~-7$. Particularly, this weight system with $w_8=-\frac{7}{2}$ satisfies both the {\it typical} and {\it real} properties.  By using these properties, we have already shown that the typical rep $(0~0~0~0~0~0~0~-\frac{7}{2})$ is b
eautifully
identified with the supermultiplets of the {\it D}=4, {\it N}=8 supergravity and {\it D}=10, {\it N}=2 chiral supergravity.$^{12,13}$ 

Now, let us consider the case of {\it D}=10, {\it N}=2 maximal chiral supergravity.  Although the hidden symmetry of full theory on the shell is still not known, we have found that SO(8)$\otimes$ SO(2) $\subset$ SU(8/1).  Here, we have introduced a bigger
 symmetry SU(8) $\supset$ SO(8) to preserve chirality, and the U(1) $\approx$ SO(2), which corresponds to the simple odd root, for {\it N}=2 supersymmetry.

Let $h_i (i=1,2,\ldots ,8)$ be Cartan sublagebras of SU(8/1) superalgebra.  Then the U(1) sublagebra is composed of $\sum_{i=1}^8 ih^i$ to satisfy the supertraceless condition, and the U(1) supercharge generator should be 
$Diag (1,1,1,1,1,1,1,8)$. Then, the typical lowest dimensional rep $(0~\cdots ~0~-\frac{7}{2})~=~
[{\bf 128}_B \oplus {\bf 128}_F]$ corresponds to the supermultiplets of
$D=10, ~N=2$ chiral supergravity.  The full contents of the representation and the field identifications are given by

\begin{equation}
\begin{array}{lccl}
\mbox{floor}     & \mbox{SU(8/1)} & \mbox{SO(8)$\otimes$U(1)} & \mbox{field} \\
\noalign{\vskip3pt}
\noalign{\hrule}
\noalign{\vskip3pt}
\\
\mid \mbox{gnd}> & (0~0~0~0~0~0~0~-\frac{7}{2})  & (0~0~0~0)(4) & \phi \\
\\
\mid \mbox{1st}> & (0~0~0~0~0~0~1~-\frac{7}{2})  & (0~0~1~0)(3) & \lambda \\
\\
\mid \mbox{2nd}> & (0~0~0~0~0~1~0~-\frac{5}{2})  & (0~1~0~0)(2) & A_{\mu \nu}\\
\\
\mid \mbox{3rd}> & (0~0~0~0~1~0~0~-\frac{3}{2})  & (1~0~0~1)(1) & \Psi_{\mu} \\
\\
\mid \mbox{4th}> & (0~0~0~1~0~0~0~-\frac{1}{2}) & (2~0~0~0)(0) & e_{\mu}^a \\
                 &                              & (0~0~0~2)(0) & A_{\mu \nu \rho \sigma} 
\\
\\
\mid \mbox{5th}> & (0~0~1~0~0~0~0~+\frac{1}{2})  & (1~0~0~1)(-1) & \overline{\Psi}_{\mu}\\
\\
\mid \mbox{6th}>   & (0~1~0~0~0~0~0~+\frac{3}{2})   & (0~1~0~0)(-2)   & \overline{A}_{\mu 
\nu}\\
\\
\mid \mbox{7th}> & (1~0~0~0~0~0~0~+\frac{5}{2})  & (0~0~1~0)(-3) & \overline\lambda\\
\\
\mid \mbox{8th}> & (0~0~0~0~0~0~0~+\frac{7}{2})  & (0~0~0~0)(-4) & \overline\phi.
\end{array}
\end{equation}
\\
\\
\noindent
It is interesting to note that this rep can be identified with a single scalar superfield $\Phi (x, \theta )$ treated by Green and Schwarz.$^{10}$

\vspace{1cm}
\begin{center}
{\large \bf III. Possible Superalgebraic Truncations}
\end{center}

\begin{center}
{\large \bf 3.1 {\it D}=8, {\it N}=1 Reduction}
\end{center}

Now, let us consider the possible superalgebraic trunction to the eight dimensions.  The supermultiplets of $D=8,~N=2$ are in the rep space of SO(6) $\otimes$ Sp(2) symmetry.  However, the irreps of SO(6) $\otimes$ Sp(2) are not fit in SU(8) $\otimes$ U(1
) $\subset$ SU(8/1).  Therefore, the case of the only possible superalgebraic truncation is to accommodate the supermultiplets of $D=8,~N=1$ in the rep space of a maximal subalgebra SU(4) $\otimes$ SU(4/1) $\otimes$ U$^{\it a}$(1), which is simply obtaine
d by removing the fourth node from the Kac-Dynkin diagram in Eq.(1). The supertraceless condition is satisfied by taking the U$^{\it a}$(1) assignment as $Diag.(3,3,3,3,-4,-4,-4,-4,-4)$.  Note that the U$^{\it a}$(1) subalgebra is composed of $[3h^1 + 6h^
2 + 9h^3 + 12h^4 + 8h^5 + 4h^6 - 4h^8]$.

We find that this branching scheme describes the $D=8,~N=1$ chiral theory.  The light-like symmetry SO(6) $\approx$ SU(4) is realized through the subalgebraic chains as follows

\begin{equation}
\begin{array}{lcl}
\mbox{SU}(8/1) & \longrightarrow &~ \mbox{SU}_{V} (4) \otimes
\mbox{SU}_{S} (4/1) \otimes \mbox{U}^{a} (1)\\
\\
        & \longrightarrow &~ \mbox{SU}_{V} (4) \otimes \mbox{SU}_{S} (4)
        \otimes \mbox{U}^{a} (1) \otimes \mbox{U}^{b} (1)\\
\\
        & \longrightarrow &~ \mbox{SU}_{V+S} (4) \otimes \mbox{U}(1),
\end{array}
\end{equation}
\\
\noindent
where the subscripts {\it V} and {\it S} mean vectorial and spinorial reps, respectively.
A branching rule of the first step for the rep $(0~0~0~0~0~0~0~-\frac{7}{2})$
is

\begin{equation}
\begin{array}{ll}
(0~0~0~0~0~0~0~-\frac{7}{2})
\longrightarrow &~(0~0~0)(0~0~0~-\frac{7}{2})(2) \oplus (0~0~1)(0~0~0~-\frac{5}{2})(1)\\
\\
     & \oplus ~(0~1~0)(0~0~0~-\frac{3}{2})(0) \oplus (1~0~0)(0~0~0~-\frac{1}{2})(-1)\\
\\
     & \oplus ~(0~0~0)(0~0~0~\frac{1}{2})(-2),\\
\end{array}
\end{equation}
\\

\noindent
where the U$^{\it a}$(1) supercharges are normalized by -7.  The typical rep
$(0~0~0~w_8)$ of SU$_{S}$(4/1) has the content of ~$({\bf 8}_B ~ + ~{\bf 8}_F )$ =
$({\bf 1} + {\bf {\bar 4}} + {\bf 6} + {\bf 4} + {\bf 1})$.

Then, the typical rep $(0~0~0)(0~0~0~-\frac{7}{2})(2)$ in Eq.(10) gives a Yang-Mills multiplet such as

\begin{equation}
\begin{array}{ccc}
\mbox{SU}_V (4) \otimes \mbox{SU}_S (4) \otimes \mbox{U}^a (1) \otimes
\mbox{U}^b (1)
& \mbox{SU}_{V+S} (4) \otimes \mbox{U}(1)   & \mbox{field} \\
\noalign{\vskip3pt}
\noalign{\hrule}
\noalign{\vskip3pt}
\\
(0~0~0)(0~0~0)(2)(-14) & (0~0~0)(-2) & \phi^1 \\
\\
(0~0~0)(0~0~1)(2)(-11)  & (0~0~1)(-1) & \chi^-\\
\\
(0~0~0)(0~1~0)(2)(-8)  & (0~1~0)(0) & A_{\mu} \\
\\
(0~0~0)(1~0~0)(2)(-5)  & (1~0~0)(+1) & \chi^+ \\
\\
(0~0~0)(0~0~0)(2)(-2)  & (0~0~0)(+2) & \phi^2.
\end{array}
\end{equation}
\\

\noindent
Here, the U(1) supercharge is given by $\mbox{U}(1)~=~\frac{1}{3} [4\mbox{U}^{a} (1)~ +~ \mbox{U}^{b} (1)]$.  Note that since the rep $(0~0~0)(0~0~0~\frac{1}{2})(-2)$ is the complex conjugation of the rep given by Eq.(11), one may also take it as a Yang-M
ills multiplet.

On the other hand, the graviton multiplet is the rep $(0~1~0)
(0~0~0~-\frac{3}{2})(0)$ in Eq.(10) as follows

\begin{equation}
\begin{array}{ccc}
\mbox{SU}_V (4) \otimes \mbox{SU}_S (4) \otimes \mbox{U}^a (1)
\otimes \mbox{U}^b (1)
& \mbox{SU}_{V+S} (4) \otimes \mbox{U}(1)   & \mbox{field} \\
\noalign{\vskip3pt}
\noalign{\hrule}
\noalign{\vskip3pt}
\\
(0~1~0)(0~0~0)(0)(-6) & (0~1~0)(-2) & A_{\mu}^1 \\
\\
(0~1~0)(0~0~1)(0)(-3)  & (0~1~1)(-1) & \Psi_{\mu}^- \\
                       & (1~0~0)(-1) & \chi^- \\
\\
(0~1~0)(0~1~0)(0)(0)  & (0~2~0)(0) & e_{\mu}^a \\
                      & (1~0~1)(0) & B_{\mu \nu}\\
                      & (0~0~0)(0) & \phi\\
\\
(0~1~0)(1~0~0)(0)(+3)  & (1~1~0)(+1) & \Psi_{\mu}^+ \\
                      & (0~0~1)(+1)  & \chi^+ \\
\\
(0~1~0)(0~0~0)(0)(+6)  & (0~1~0)(+2) & A_{\mu}^2 .\\
\end{array}
\end{equation}
\\
\noindent
Note that the other two reps [$(0~0~1)(0~0~0~-\frac{5}{2})(1) \oplus (1~0~0)(0~0~0~-\frac{1}{2})(-1)$] in Eq.(10) make an extra gravitino multiplet at the $\mbox{SU}_{V+S} (4) \otimes \mbox{U}(1)$ stage, which should be removed for consistency in the $D=8
,~N=1$ theory.

\vspace{1cm}
\begin{center}
{\large \bf 3.2 {\it D}=6, {\it N}=2 Reduction}
\end{center}

As you know, the underlying symmetry of $D=6,~ (N_+ ,~N_-)$ = (2, 0) chiral theory is SO(4) $\otimes$ Sp(2).  But, let us try to accomodate this symmetry in the larger supersymmetry SU(4/1), which contains SU(2) $\otimes$ SU(2/1) substructure given by the
 branching pattern

\begin{equation}
\begin{array}{lcl}
\mbox{SU}_{V} (4) \otimes 
\mbox{SU}_{S} (4/1) & \longrightarrow &~ \mbox{Sp}_{V} (4)       \otimes \mbox{SU}_{S} (2) \otimes \mbox{SU}_{S} (2/1) \\
\\
                    & \longrightarrow &~ [\mbox{SU}_{V} (2)]^2 \otimes [\mbox{SU}_{S} (2)]^2\\
\\
                    & \longrightarrow &~ [\mbox{SU}_{V+S} (2)]^2 ~\approx ~\mbox{SO}_{V+S} (4).
\end{array}
\end{equation}
\\

\noindent
Here, we use the branching pattern  SU$_V$(4) $\longrightarrow $ 
Sp$_V$(4) $\longrightarrow$ SU$_V(2) \otimes$ SU$_V$(2) for the vectorial space.  Thus, the branching rule for the rep $(0~1~0)$ of SU$_V$(4) should be $(0~1~0)~\longrightarrow$ $(0~1) \oplus (0~0)$ $\longrightarrow$ $[(1)(1) \oplus (0)(0)] \oplus (0)(0)$
.  Note that we have lost the U(1) symmetry at the 
$\mbox{Sp}_{V} (4)$ branching stage. On the other hand, we use the branching pattern $\mbox{SU}_{S} (4/1) ~\longrightarrow ~ \mbox{SU}_{S} (2) \otimes \mbox{SU}_{S} (2/1)$ for the spinorial space.  Then, the rep $(0~0~0~-3/2)$ of $\mbox{SU}_{S} (4/1)$ bra
nches into the reps $(0)(0~ -3/2) 
\oplus (1)(0~-1/2) \oplus (0)(0~1/2)$.    Finally the rep [(1)(1)][(1)(0 -1/2)] in $[\mbox{SU}_{V} (2)]^2 \otimes [\mbox{SU}_{S} (2)\otimes \mbox{SU}_{S} (2/1)]$ basis 
reduces to the following reps of $\mbox{SO}_{V+S} (4)$

\begin{equation}
[(2~1) \oplus (0~1)] \oplus [(2~2) \oplus (0~2) \oplus (2~1) \oplus (0~0)] \oplus [(2~1) \oplus (0~1)],
\end{equation}
\\
\noindent
where the reps are denoted in Dynkin weights of SO(4) and the floors of SU(2/1) are distinguished by the square brackets. Note that they are equivalent to the following expression of Strathdee$^5$

\begin{equation}
(3,2;1) \otimes 2^2 \oplus (1,2;1) \otimes 2^2,
\end{equation}
\\
\noindent
which are the reps of SO(4) $\otimes$ Sp(2). Here, the rep $2^2$ means (1,2;1) $\oplus$ (1,1;2) in 
SO(4) $\otimes$ Sp(2) basis.
On the other hand, the rep $(0)(0~-3/2)$ of $\mbox{SU}(2) \otimes \mbox{SU}(2/1)$ is 

\begin{equation}
\begin{array}{lc}
                    & (0)(0~-\frac{3}{2})\\
\\
\mid \mbox{ground}> & (0)(0~-\frac{3}{2})\\
\\
\mid \mbox{1st}> & (0)(1~-\frac{3}{2})\\
\\
\mid \mbox{2nd}> & (0)(0~-\frac{1}{2}),\\
\end{array}
\end{equation}
\\

\noindent
where the first floor gives (1,2;1) and the ground and the second floors make (1,1;2), that is,
the Sp(2) indices are reproduced by the floors.

Similarly, the Yang-Mills and matter multiplets can be also easily identified with the reps
$(0)_V (0)_V (1)_S (0 -1/2)_S$ and $(0)_V (0)_V (0)_S (0 -3/2)_S$, respectively. Note that the adjoint rep of SU(2/1) is given by

\begin{equation}
\begin{array}{lccc}
                    & (1~-1)\\
\\
\mid \mbox{ground}> & (1~-1) & = & Q_{1/2}^+ \\
\\
\mid \mbox{1st}> & (2~-1) & = & \mbox{SU(2)} \\
                 & (0~0) & = & \mbox{U(1)} \\
\\
\mid \mbox{2nd}> & (1~0) & = & Q_{1/2}^-. \\
\end{array}
\end{equation}

\noindent
As a results, the Yang-Mills multiplet $(0~0~0)(0~0~0~-\frac{7}{2})(2)$ in $D=8, ~N=1$
reduces into $(0)(0~-\frac{7}{2})~\oplus~(1)(0~-\frac{5}{2})~\oplus~
(0)(0~-\frac{3}{2})$ in $D=6, ~N=2$. The Yang-Mills multiplet of $D=6,~N=2$ is
$(1)(0~-\frac{5}{2})=[\bf{(2,1) \oplus (2,2) \oplus (2,1)}]$, while the other
reps $(0)(0~w_8)=[\bf{(1,1) \oplus (1,2) \oplus (1,1)}]$ are matter multiplets.

It seems appropriate to comment on the truncations from {\it D}=6 to {\it D}=4 theories. 
Unfortunately, we cannot directly obtain the {\it D}=4, {\it N}=4, 3, 2, 1 from {\it D}=6, {\it N}=2 
theory by the successive superalgebraic truncations because we already lost 
several U(1) informations at the {\it D}=6 stage. 
However, the {\it D}=4, {\it N}=8 supergravity is effectively equivalent to the {\it D}=10,
{\it N}=2 chiral supergravity, and these equivalence can be shown schematically in the SU(8/1) Kac-Dynkin diagram

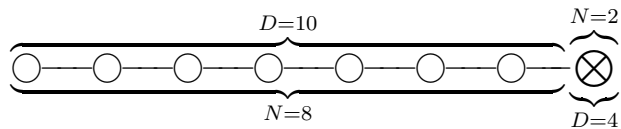
\begin{eqnarray}
\underbrace{\overbrace{
\bigcirc \!\!-\!\!\!-\!\!\!-\!\!\bigcirc \!\!-\!\!\!-\!\!\!-\!\!
\bigcirc \!\!-\!\!\!-\!\!\!-\!\!\bigcirc \!\!-\!\!\!-\!\!\!-\!\!
\bigcirc \!\!-\!\!\!-\!\!\!-\!\!\bigcirc \!\!-\!\!\!-\!\!\!-\!\!
\bigcirc \!\!-\!\!\!-\!\!\!-\!\!}^{D=10}}_{N=8}
\underbrace{\overbrace{ \bigotimes}^{N=2}}_{D=4} 
\nonumber
\end{eqnarray}
\\
\\
\noindent
In fact, this equivalence  implies that  as space-time   dimensions are decreased  by the   consistent dimensional reduction, supersymmetry must be extended or vice versa.$^{16}$  According to this line, it seems enough to comment our previous result$^{13
}$ that  the successive superalgebraic truncations 
from {\it D}=4, {\it N}=8 theory to  {\it D}=4, {\it N}=7,6, ${\cdots}$, 1 supergravity  theories can be
systematically realized as sub-superalgebraic chains of SU(8/1) superalgebra.

\vspace{1cm}
\begin{center}
{\large \bf IV. Conclusion}
\end{center}

In conclusion, we have studied {\it D}=10, {\it N}=2 chiral supergravity in the context of SU(8/1) superalgebra.  We have obtained possible regular maximal branching patterns in terms of Kac-Dynkin weight techniques.  Then, we have shown that the possible
 superalgebraic trunctions from the {\it D}=10, {\it N}=2 maximal chiral theory to the {\it D}=8, {\it N}=1, and $D=6,~N=2$ theories can be systematically realized as sub-superalgebra chains of the SU(8/1) superalgebra.
As results, we have explicitly identified the supermultiplets of the possible relevant lower dimensional theories, which have been classified in terms of super-Poincar\'{e} algebra by Strathdee, with irreps of SU({\it N}/1) superalgebra by using the syste
matic superalgebraic truncation method.  Finally, through further investigations, we hope that our superalgbraic branching method will provide a deeper understanding of
the structure of the supersymmetric systems including the M and F theories.

\vspace{1cm}

\begin{center}
{\bf Acknowledgements}
\end{center}

The present study was supported by the Basic Science Research
Institute Program,
Ministry of Education, 1996, Project No. {\bf BSRI}-2414.

\newpage

\begin{center}
{\bf REFERENCES}
\end{center}

\begin{description}
\item{1.} V. Kac, Adv. in Math. {\bf 26}, 8 (1977); Commun. Math. Phys.
{\bf 53}, 31 (1977).
\item{2.} Y.A. Gol'fand and E.P. Likhtman, Pis'ma Zh. Eksp. Theor. Fiz.
 {\bf 13}, 452(1971) [JETP Lett. {\bf 13},323(1971)];  A. Neveu and J. H. Schwarz, Nucl. Phys. {\bf B31}, 86 (1971); P. Ramond, Phys. Rev.
 {\bf D3}, 2415 (1971); P. G. O. Freund and  I. Kaplansky, J. Math. Phys.
 {\bf 17}, 228 (1976); S. Deser and B. Zumino, Phys. Lett. {\bf 62B}, 335 (1976);
D. Z. Freedman, P. van Nieuwenhuizen, and S. Ferrara, Phys. Rev. {\bf D13},
3214 (1976); Y. Ne'eman, Phys. Lett. {\bf 81B}, 190 (1979); D. B. Fairlie,
 {\it ibid.} {\bf 82B}, 97 (1979);  F. Iachello, Phys. Rev. Lett. {\bf 44}, 772 (1980); A.B. Balantekin,I. Bars, and F. Iachello, {\it ibid.} {\bf 47}, 19 (1981);   P. van Nieuwenhuizen, Phys. Rep. {\bf 68}, 189 (1981); M. B. Green and J. H. Schwarz, Nucl
. Phys. {\bf B198}, 474 (1982); J. P. Hurni and B. Morel, J. Math. Phys. {\bf 24},
 157 (1983).
\item{3.} J. Schwarz, Phys. Lett. {\bf B367}, 97 (1996); C. Vafa, "Evidence for F-theory",
hep-th/9602022.
\item{4.} I. Bars, Phys. Rev. {\bf D54}, 5203 (1996).
\item{5.} J. Strathdee, Int. J. Mod. Phys. {\bf A2}, 173 (1987).
\item{6.} E. Cremmer, in {\it Superspace and Supergravity}, edited by
S. Hawking and M. Rocek (Cambridge University Press, London, England, 1980).
\item{7.} K. Y. Kim, C. H. Kim, Y. Kim, and Y. J. Park,
J. Korean Phys. Soc. {\bf 18}, 249 (1985);  C. H. Kim,
K. Y. Kim, W. S. l'Yi, Y. Kim, and Y. J. Park, Mod. Phys. Lett. {\bf A3},
1005 (1988); C. H. Kim, Y. J. Park, K. Y. Kim, Y. Kim, and W. S. l'Yi, Phys.
Rev. {\bf D44}, 3169 (1991).
\item{8.} C. H. Kim, K. Y. Kim, Y. Kim, H. W. Lee, W. S. l'Yi, and
Y. J. Park, Phys. Rev. {\bf D40}, 1969 (1989).
\item{9.} C. H. Kim, K. Y. Kim, W. S. l'Yi, Y. Kim, and Y. J. Park,
J. Math. Phys. {\bf 27}, 2009 (1986).
\item{10.} M. B. Green and J. H. Schwarz, Phys. Lett. {\bf 122B}, 143 (1983).
\item{11.} J. H. Schwarz and P. C. West, Phys. Lett. {\bf 126B}, 301 (1983);
P. Howe and P. C. West, Nucl. Phys. {\bf B238}, 181 (1984).
\item{12.} C. H. Kim, K. Y. Kim, Y. Kim, and Y. J. Park, Phys.
Rev. {\bf D39}, 2967 (1989); C. H. Kim, S. Cho, S. H. Yoon, and Y. J. Park,
J. Korean Phys. Soc. {\bf 26}, 603 (1993).
\item{13.} C. H. Kim, Y. J. Park, and Y. Kim, Mod. Phys. Lett. {\bf A10}, 1929 (1995).
\item{14.} R. Slansky, Phys. Rep. {\bf 79}, 1 (1981); C. H. Kim, Y. J. Park,
I. G. Koh, K. Y. Kim, and Y. Kim, Phys. Rev. {\bf D27}, 1932 (1983).
\item{15.} C. H. Kim, Y. J. Park, K. Y. Kim, and Y. Kim,
J. Korean Phys. Soc. {\bf 25}, 87 (1992).
\item{16.} G. Parisi and N. Sourlas, Phys. Rev. Lett. {\bf 43}, 744 (1979).

\end{description}
\end{document}